\documentclass[aps,preprint]{revtex4}
 
\newcommand{\eexp}{{\rm e}} 
 
\newcommand{\Dt}{{\Delta t}} 
\newcommand{\NT}{{N_{\rm T}}}
\newcommand{\Stild}{\tilde{S}} 
 \newcommand{\Rtild}{{\tilde R}}
 \newcommand{\Rbar}{{\overline R}_0} 
\newcommand{\Itild}{{\tilde I}}

\usepackage{epsfig,graphicx,bm,amssymb,amsmath}
\begin{document}  

\title{The first 100 days:
modeling the evolution of the COVID-19 pandemic
}
\author{Efthimios Kaxiras, George Neofotistos, Eleni Angelaki
}
\affiliation{Institute for Applied Computational Science, 
Harvard J.A. Paulson School of Engineering 
and Applied Sciences, Harvard University, Cambridge, MA, USA
}

\date{\today}

\begin{abstract}
A simple analytical model for modeling the evolution of the 2020 COVID-19 pandemic is presented. 
The model is based on the numerical solution of the widely used Susceptible-Infectious-Removed (SIR) populations 
model for describing epidemics.  We consider an expanded version of the original Kermack-McKendrick model, 
which includes a decaying value of the parameter $\beta$ (the effective contact rate) due to 
externally imposed conditions, to which we refer as the forced-SIR (FSIR) model. 
We introduce an approximate analytical solution to the differential equations that represent the FSIR model 
which gives very reasonable fits to real data for a number of countries over a period of 100 days (from the first onset of exponential increase, in China). 
The proposed model contains 3 adjustable parameters which are obtained by fitting actual data (up to April 28, 2020).  
We analyze these results to infer the physical meaning of the parameters involved. 
We use the model to make predictions about the total expected number of infections in each country
as well as the date when the number of infections will have reached 99\% of this total. We also compare key findings of the model with recently reported results on the high contagiousness and rapid spread of the disease.
\end{abstract}

\maketitle


The recent pandemic due to the COVID-19 virus has created unprecedented turmoil 
and changed the daily life of people over the entire planet.  It has also yielded 
a grim toll of victims that succumb to its attack. 
While there is great expertise in the medical community and the community of 
statisticians in dealing with epidemics, less is known about this particular disease 
to make reliable predictions for the evolution of the current pandemic. 

In studying past epidemics, scientists have systematically applied ``random mixing'' 
models which assume that an infectious individual may spread the disease to any susceptible member of the population, 
as originally considered by Kermack and McKendrick~\cite{Kermack_1927}.
More recent modeling approaches considered  
contact networks in which the epidemic spreads only across the edges of a contact network within a population ~(\cite{Barthelemy_2005} \cite{Ferrari_2006} \cite{Volz_2008}), 
Bayesian inference models \cite{Groendyke_2011}, models of spatial contacts in large-scale artificial cities~\cite{Zhang_2016}, 
and computational predictions of protein structures~\cite{Jumper_2020}, to name just a few of the modeling efforts. 

In the case of COVID-19, there is considerable uncertainty in the data collected about infected individuals 
due to the difficulty of testing large numbers of suspected cases. 
Although a avalanche of research studies are currently investigating the COVID-19 epidemiological characteristics ~(\cite{Sanche_2020} 
\cite{Li_2020} \cite{Imai_2020} \cite{Rothe_2020} \cite{Wynants_2020}), 
it appears that a simple model which can capture the basic behavior of the 
pandemic phenomenon, in spite of the large uncertainty in the data, can 
possibly offer useful guidance for its near-term and longer-term evolution. 
This paper aims to provide such a simple model with very few adjustable parameters.

\section{The model}

\subsection{Derivation of the model}

The original mathematical description of the spread of an infectious  
disease in a population is the so-called SIR model, due to Kermack and McKendrick~\cite{Kermack_1927} 
which divides the (fixed) population of $N$ individuals into three groups:
\begin{itemize}
\item
$S(t)$ the number of individuals susceptible but not yet infected with the disease;
\item
$I(t)$ the number of infected individuals;
\item
$R(t)$ the number of individuals removed (recovered) from the infected group, either by becoming healthy again with long-term immunity or by passing away.
\end{itemize}

The SIR model involves two positive parameters, $\beta$ and $\gamma$ which have the following 
meaning:\\
- $\beta$ describes the effective contact rate of the disease: 
an infected individual comes into contact with $\beta$ other individuals per unit time 
(the fraction that are susceptible to contracting the disease is $S/N$); \\
- $\gamma$ is the mean removal (recovery) rate, that is, $\frac{1}{\gamma}$ is the mean 
period of time during which an infected individual can pass it on before being removed from the group of the infected individuals.\\\\
This model obeys the following differential equations:
\begin{subequations}
\label{eq:SIR}
\begin{eqnarray}
\frac{dI}{dt} = \beta I \frac{S}{N} - \gamma I  
\label{eq:SIR_I}
\\
\frac{dS}{dt} = - \beta I \frac{S}{N} 
\label{eq:SIR_S}
\\
\frac{dR}{dt} = \gamma I 
\label{eq:SIR_R}
\end{eqnarray}
\end{subequations}

Many recent studies have attempted to model the data of the COVID-19 pandemic 
by imposing time-dependence conditions on the rates $\beta$ and $\gamma$ involved in the original SIR 
model, in order to account for the imposition of social-distancing measures, quarantine of infected individuals, 
and other interventions designed to slow down the spread of the disease. 
Motivated by such considerations, 
we will introduce a variation of the original model in which the parameter $\beta$ is a time-dependent, 
{\em monotonically decreasing} function.  This change 
can drastically affect the evolution of the populations.  We give below a specific example to illustrate this point.
Since the presence of time-dependence in $\beta$ introduces a forcing term, which for reasonable parameter 
values lowers the number of infected individuals (``flattens the curve'').

\begin{figure}[h]
\hspace{-0.04\textwidth}
\includegraphics[width=0.52\textwidth]{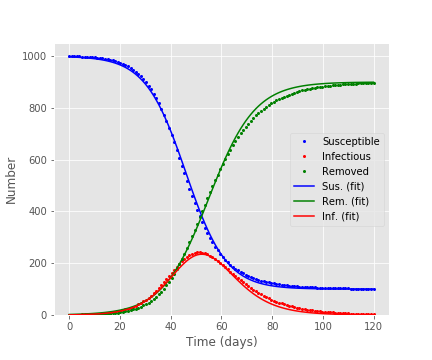}
\hspace{-0.04\textwidth}
\includegraphics[width=0.52\textwidth]{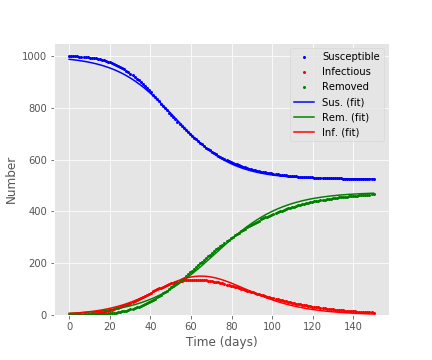}
\caption{
\small{
Numerical solution of the models, giving the susceptible $S(t)$ (blue points), 
infected $I(t)$ (red points)
and removed $R(t)$ (green points) populations as functions of time $t$ in days;
the corresponding colored lines give the approximate solutions obtained by 
the analytical expressions, Eq.~(\ref{eq:FSIR}). 
{\bf Left}: The SIR model, with parameter values $\beta=0.25$ and $\gamma = 1/10$; 
the parameter $\beta$ is constant.  
{\bf Right}: The SIR model with a time-dependent parameter $\beta$ with exponential decay and
parameter values $\beta_0=0.25$, $\gamma = 1/20$, $\lambda=50$
(see text for details).  \\
}
}
\label{fig:FSIR_model}
\end{figure}

The system of equations that describe the SIR, with or without the time-dependence in the
parameter $\beta$,
can be easily solved numerically, as shown in Fig. \ref{fig:FSIR_model}, 
giving the three group populations ($S, I, R$) as a function of time. 
Kermack and McKendrick pointed out the ``it is impossible from these equations 
to obtain $I(t)$ as an explicit function of $t$'' (p. 713, \cite{Kermack_1927}), 
but provided approximations valid under certain conditions. 
Here we aim to give a simple approximate analytical solution inspired by the numerical solution.

We observe from the numerical solution shown of the SIR model, shown in Fig. \ref{fig:FSIR_model},
that both the susceptible and the removed populations ($S$ and $R$, respectively) 
behave like sigmoids, which is the typical behavior of solutions to 
differential equations that involve exponential growth and decay. 
Moreover, the infected population is always given by the following expression
\begin{equation}
I(t) = N - S(t) - R(t)
\label{eq:infect-S-R}
\end{equation}
From these observations, we take the approximate solutions to be given by:
\begin{subequations}
\label{eq:FSIR}
\begin{eqnarray}
\Stild & = & N - \frac{N'}{1+\eexp^{-\alpha_1(t-t_1)}} 
\label{eq:FSIR_S}
\\
\Rtild & = & \frac{N'}{1+\eexp^{-\alpha_2(t-t_2)}} 
\label{eq:FSIR_R}
\\
\Itild & = & N - \Stild(t) - \Rtild(t) =  \frac{N'}{1+\eexp^{-\alpha_1(t-t_1)}}
- \frac{N'}{1+\eexp^{-\alpha_2(t-t_2)}} 
\label{eq:FSIR_I}
\end{eqnarray}
\end{subequations}
where $N'$, $\alpha_1, \alpha_2$, $t_1, t_2$ are treated as adjustable 
parameters, with $t_1$  and $t_2$ representing the times at which the $\Stild$ and $\Rtild$ 
populations reach their sigmoid midpoint values, respectively.
Interestingly, the analytical expressions introduced above fit even better the numerical solution 
of the model with a time-dependent $\beta$ parameter. 
In Fig. \ref{fig:FSIR_model} we give examples of how well the 
approximate analytical expressions fit the ``exact'' numerical ones. In these examples,
for the model with time-dependent $\beta$ we assumed $\beta(t) = \beta_0 \exp(-t/\lambda)$, 
although we emphasize that this assumption is for illustrative purposes only and does not 
affect the general behavior of the model.  
Indeed, as we show below, $\beta(t)$ has the behavior of a sigmoid. 
For the SIR model in the example of Fig. \ref{fig:FSIR_model}, 
the fit to the analytical expression of Eq.~(\ref{eq:FSIR}) has an RMSE value of $9.4$ and the integral 
of the $I(t)$ values differs from the exact result by $-9.5$\%.
For the model with time-dependent $\beta$ in the example we considered, 
the fit to the analytical model of Eq.~(\ref{eq:FSIR}) has an RMSE value of $8.3$ and 
the integral of $I(t)$ differs from the exact result by $0.3$\%. 

Since the analytical model of Eq.~(\ref{eq:FSIR}) can capture the behavior of the SIR model 
including a time-dependent $\beta$, which represent the ``forcing'' or ``flattening'' of the 
curve of infected individuals, we refer to it as the ``FSIR'' model.  

\subsection{Analysis of the model}

Here we derive relations between the parameters used in the model of 
Eq.~(\ref{eq:FSIR}), and the parameters of the original set of differential 
equations, Eq.~(\ref{eq:SIR}). 
To keep the expressions simple, we will assume $\alpha_1=\alpha_2=\alpha$ 
and define $\Delta t = t_2 - t_1$. 
By inserting the expressions for $\Stild$ and $\Itild$ in Eq.~(\ref{eq:SIR_S}) we find: 
\begin{equation}
\beta(t) = \frac{\alpha}{\eexp^{\alpha \Delta t} - 1} \left ( 
\frac{1+\eexp^{-\alpha(t-t_1)} \eexp^{\alpha \Delta t}}{1-n'+\eexp^{-\alpha(t-t_1)}} \right )
\label{eq:beta}
\end{equation}
where we have defined $n'=N'/N$. 
Similarly, by inserting the expressions for $\Rtild$ and $\Itild$ in Eq.~(\ref{eq:SIR_R}) we obtain:
\begin{equation}
\gamma(t) = \frac{\alpha}{1-\eexp^{-\alpha \Delta t}} \left ( 
\frac{1+\eexp^{-\alpha(t-t_1)}}{1+\eexp^{-\alpha(t-t_1)}\eexp^{\alpha \Delta t}} \right )
\label{eq:gamma}
\end{equation}
Thus, in the approximate model described by Eq.~(\ref{eq:FSIR}), 
the parameters $\beta$ and $\gamma$ of the original SIR model become time-dependent, 
if we treat $\alpha$ as constant to be determined by fitting the data (see next section). 
In the FSIR model the effect of interventions and measures can 
be inferred from the values of the adjustable parameters $t_1$, $\Dt$ 
and $N'$, as will be explained in the next section, so that 
there is no need to impose specific time-dependent conditions on the model parameters themselves.

The quantity $n'=N'/N$ we defined in the expression of $\beta(t)$ is the fraction of the original 
susceptible population that was infected, and thus does become part of the removed population. 
There are two possible limiting values for this quantity: $n'\to 1$, the limit in which the 
entire susceptible population was exposed and eventually becomes removed population, and 
$n' \to 0$, the limit in which only a tiny fraction of the susceptible population was exposed.
In the first limit we obtain:
\begin{equation}
n' \to 1 \Rightarrow 
\beta_1(t) = \frac{\alpha}{\eexp^{\alpha \Delta t} - 1} \left ( \eexp^{\alpha(t-t_1)} + \eexp^{\alpha \Delta t}\right ) 
\end{equation}
while in the second limit we obtain:
\begin{equation}
n' \to 0 \Rightarrow 
\beta_2(t) = \frac{\alpha}{\eexp^{\alpha \Delta t} - 1} \left ( 
\frac{1+\eexp^{-\alpha(t-t_1)} \eexp^{\alpha \Delta t}}{1+\eexp^{-\alpha(t-t_1)}}\right ) 
\end{equation}
From the first expression we see that for $t\gg t_1$ the value of $\beta_1 (t)$ increases exponentially, 
which is an unphysical result. From the second expression, we see that 
$\beta$ is a {\em monotonically  decreasing} function of time and 
for $t \gg t_1$  tends to the constant value 
$$\lim_{t\gg t_1} \beta_2(t) = \alpha/(\eexp^{\alpha \Delta t} - 1),$$ 
which is the expected behavior in the FSIR model.  

For $t\ll t_1$ and assuming that $\alpha \Delta t > 1$ we find that 
$$\beta_2 \approx  \alpha \frac{\eexp^{\alpha \Delta t}}
{\eexp^{\alpha \Delta t} - 1} \approx \alpha,$$
which relates the adjustable parameter $\alpha$ of the analytical model 
to the value of the parameter $\beta$ appearing in the original SIR model.

The quantity $R_0 = \beta / \gamma$ of the SIR model 
is used to estimate the value of the basic reproduction number of an epidemic. 
From our analytical model, in the limit $n'\to 0$, the quantity $\beta/\gamma$ takes the form
\begin{equation}
\beta/\gamma = \eexp^{-\alpha \Delta t} 
 \left ( 
\frac{1+\eexp^{-\alpha(t-t_1)} \eexp^{\alpha \Delta t}}{1+\eexp^{-\alpha(t-t_1)}} \right )^2
\label{eq:R_0}
\end{equation}
For $t=0$, and assuming that $\alpha t_1\gg 1$  
(as is the case for the fits to reported data discussed in the next section),  
 this quantity becomes
$$t = 0: \; \; \;  \beta/\gamma = 
\eexp^{-\alpha \Delta t} 
 \left ( 
\frac{1+\eexp^{\alpha t_1} \eexp^{\alpha \Delta t}}{1+\eexp^{\alpha t_1}} \right )^2
\approx \eexp^{\alpha \Delta t}.$$
For $t \gg t_2$, the quantity $\beta/\gamma$ becomes
$$ t \gg t_2 : \; \; \; \beta/\gamma \approx \eexp^{-\alpha \Delta t}.$$
The first number is very large for typical values of the parameters in the FSIR model obtained 
from fits to reported data, while the second value is very small, close to zero. 
Neither result is realistic.  In the important range $t_1 < t < t_2$, 
we find from numerical results that this quantity is approximately described 
by a decaying exponential in time
$$ t_1 < t < t_2 : \; \; \; \beta/\gamma \sim \eexp^{- t/\lambda},$$
with $\lambda \approx 1/2\alpha$.
This result implies that in this range we would expect $\beta \sim  \eexp^{- \alpha t}$
(the functional form we assumed for illustrative purposes in Fig. \ref{fig:FSIR_model}), 
and $\gamma \sim  \eexp^{\alpha t}$.
From this last expression, taking the time average of $\gamma$ 
in the range $t_1 < t < t_2=t_1+\Delta t$, which we call $\overline{\gamma}$,
we find 
$$\overline{\gamma} = \frac{1}{t_2-t_1} \int_{t_1}^{t_2} \gamma_2  \eexp^{\alpha (t-t_2)} d t = 
\frac{\gamma_2}{\alpha \Delta t} \left [ 1 - \eexp^{-\alpha \Delta t} \right ] \approx \frac{1}{\Delta t}$$
where we have used
$$\gamma_2 \approx \frac{\alpha}{1-\eexp^{-\alpha \Delta t}} $$   
from the expression of Eq.~(\ref{eq:gamma}), a reasonable approximation for $t \gtrsim t_2$. 
The last relation for the average value of $\gamma$ 
is obeyed to a good approximation for each case of reported data we examined.

Using the preceding analysis that led to the relations $\beta \approx \alpha$ for the 
initial value of $\beta$ and 
$\overline{\gamma} \sim 1/\Delta t$ for the average value of $\gamma$, 
we suggest that a reasonable representation of the 
quantity $\beta/\gamma$ is given by the value of $\alpha \Delta t$.
Thus, we will use this value as a proxy for $R_0$, and will refer to it as $\Rbar$. 
The parameters estimated from the fit of our analytical model to reported data
give a value of $\Rbar$ which is in agreement with the recently reported median value
of $R_0$ for the pandemic.

\section{Application to reported data}

\begin{figure}[h]
\includegraphics[width=1.0\textwidth]{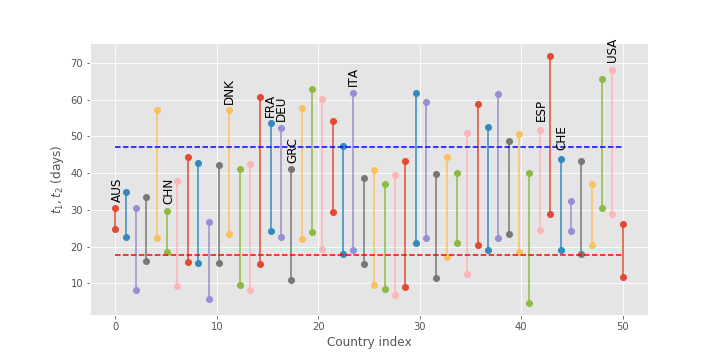}
\caption{
\small{
The values of the parameters $t_1$ and $t_2=t_1+\Delta t$ for 50 countries, 
obtained by fitting the FSIR model with data up to 28 April 2020.
The dashed lines give the average values of the parameters, $\langle t_1 \rangle$ (red) 
and $\langle t_2 \rangle = \langle t_1 \rangle + \langle \Delta t \rangle$ (blue)
of the FSIR model defined by Eq.~(\ref{eq:FSIR}).  
The countries with label are used to examine the behavior of the model in more detail.
}
}
\label{fig:50Country_stat}
\end{figure}

We use our analytical FSIR model to fit the behavior of 
infected populations of different countries, as obtained from \cite{ECDC_source}, 
for a period ending on April 28, 2020 
which corresponds to approximately 100 days from the onset of the exponential growth of reported cases in China. 

In order to obtain a meaningful fit, we had to consider data for each country that show a 
monotonic increase at the beginning.  This means that a few data points in 
each case were excluded, as they corresponded to sporadic reports 
of very few isolated cases, typically 1 to 10 in a given day, interspersed by several days of zero cases.  
In practice this means that the fitting begins at a certain cutoff day denoted as $t_0$. 
In order to make the fit more robust and simpler, we chose $\alpha_1=\alpha_2=\alpha$. 
Moreover, we found by trial-and-error that the value $\alpha=0.25$ is 
the optimal choice for all the countries we considered.  
This  leaves three adjustable parameters in the model that 
can be varied to obtain the best fit to the data, namely $t_1$, $t_2$ and $N'$; 
instead of $t_1$ and $t_2$, we elected to use instead as independent 
parameters $t_1$ and $\Delta t = t_2-t_1$. 
The best fit here is defined in the Root-Mean-Square (RMS) sense. 

We were able to obtain reasonable fits for 50 countries from the entire database~\cite{ECDC_source}.
The resulting values for the parameters $t_1, t_2$, are shown in Fig. \ref{fig:50Country_stat}
($\Delta t = t_2-t_1$ is the distance between each pair of values).
The averages and standard deviations for this set are $\langle t_1 \rangle =17.81 \pm 6.58$, 
$\langle \Delta t \rangle = 29.20 \pm 9.16$, 
giving $\langle t_2 \rangle = \langle t_1 \rangle + \langle \Delta t \rangle = 47.19$. 
The values of the parameters involved span a wide range.  For other countries in the 
database, the data are either 
too noisy or have not reached the point where the FSIR model can provide a good fit: specifically, 
the model needs to include data up to the maximum of the curve, otherwise it does not 
give meaningful values for the fitting parameters.
 
\begin{figure}[h]
\hspace{-0.04\textwidth}
\includegraphics[width=0.52\textwidth]{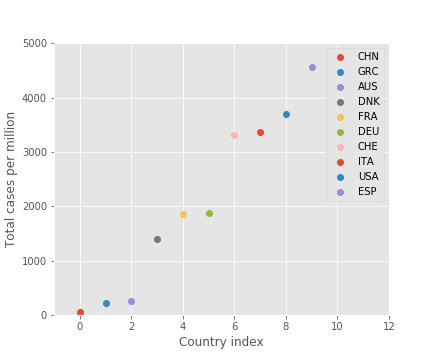}
\hspace{-0.04\textwidth}
\includegraphics[width=0.52\textwidth]{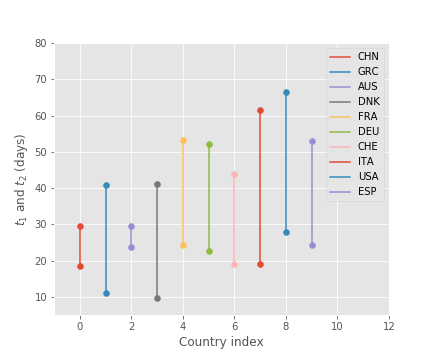}
\caption{
\small{
{\bf Left}: the estimated total cases ($N_T$) scaled by the population of each country.
{\bf Right}: 
$t_1$ and $t_2$ values (dots)  in 10 countries, 
as obtained in the FSIR model by fitting the raw data reported in \cite{ECDC_source}, 
including data up to April 28, 2020 (see also Fig. \ref{fig:data_outliers} for specific examples).  
}
}
\label{fig:daily_10}
\end{figure}

 Instead of including all 50 countries in the following discussion, we have chosen 
 to focus on 10 countries that span the whole range of parameter values and could hopefully 
 provide some insight to the behavior of the pandemic. 
The choice of the 10 countries also aims to represent parts of the world more heavily 
or less heavily impacted by the disease, as well as more typical cases.  
Here we defined the impact as the total number $N_T$ of infected individuals during the 
first wave of the pandemic, as predicted by the FSIR model; this number is 
scaled by the population of the country, $N_P$, in Fig. \ref{fig:daily_10}.
In particular, we have included 3 countries in which the impact was small, China, Greece and 
Australia for which $(N_T/N_P)<500 $ infected per million, 
3 countries in which the impact was moderate, Denmark, Germany and France for which 
$1,000< (N_T/N_P) < 2,000$ infected per million, 
and four countries where the impact was large, Switzerland, Italy, USA and Spain for which 
$(N_T/N_P) > 3,000$ infected per million.  
The average value for $t_1$ for this set of 10 countries is $\langle t_1 \rangle = 20.04$ and 
for $\Delta t$ it is  $\langle \Delta t \rangle = 27.31$.

\begin{figure}[h]
\hspace{-0.04\textwidth}
\includegraphics[width=0.52\textwidth]{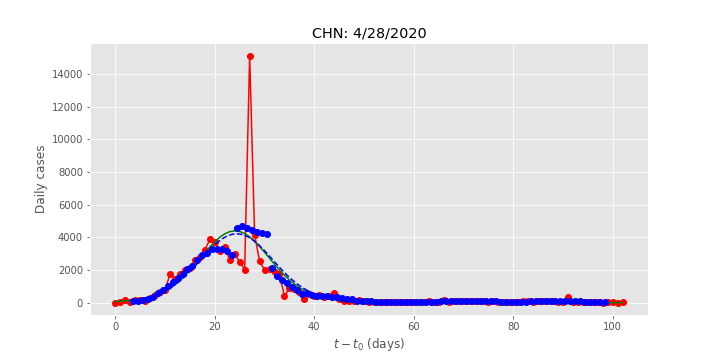}
\hspace{-0.04\textwidth}
\includegraphics[width=0.52\textwidth]{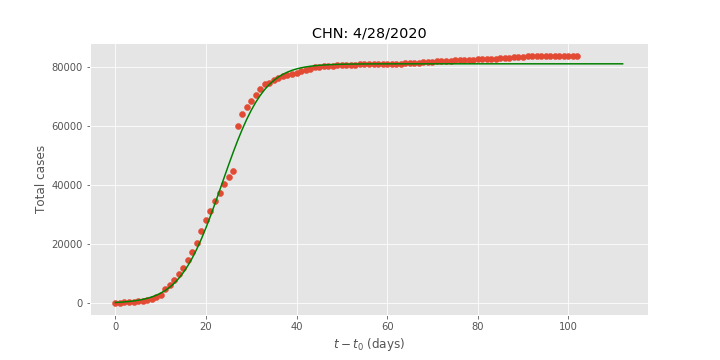}

\hspace{-0.04\textwidth}
\includegraphics[width=0.52\textwidth]{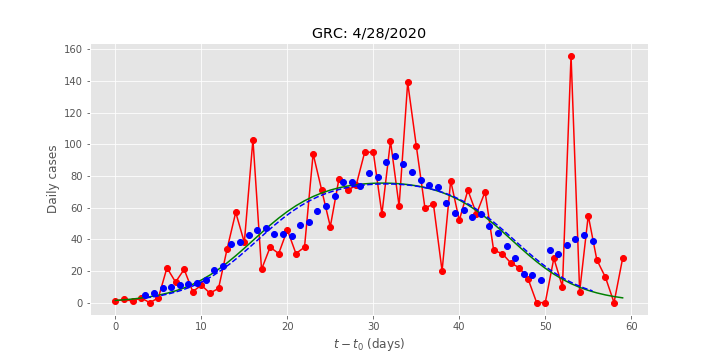}
\hspace{-0.04\textwidth}
\includegraphics[width=0.52\textwidth]{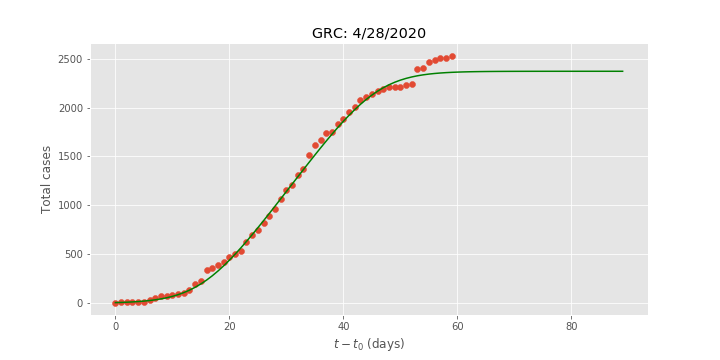}

\hspace{-0.04\textwidth}
\includegraphics[width=0.52\textwidth]{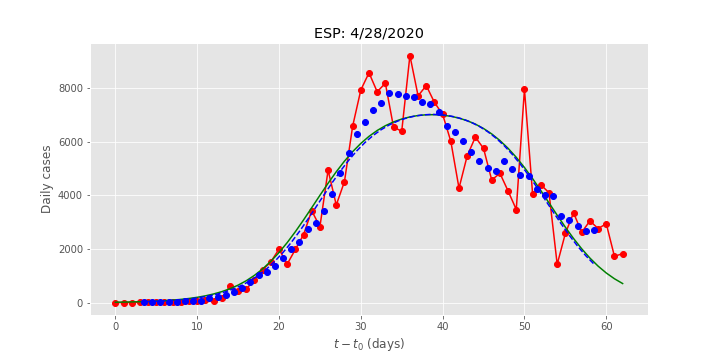}
\hspace{-0.04\textwidth}
\includegraphics[width=0.52\textwidth]{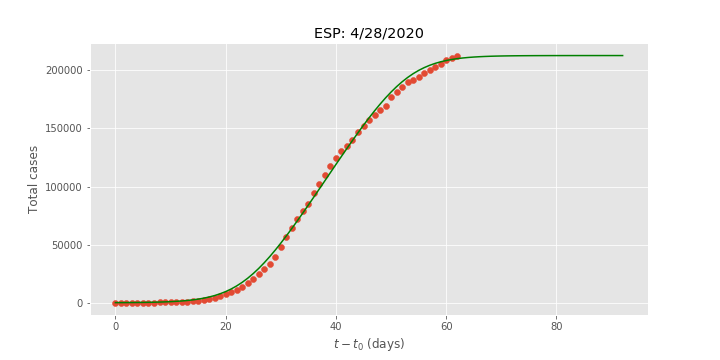}

\hspace{-0.04\textwidth}
\includegraphics[width=0.52\textwidth]{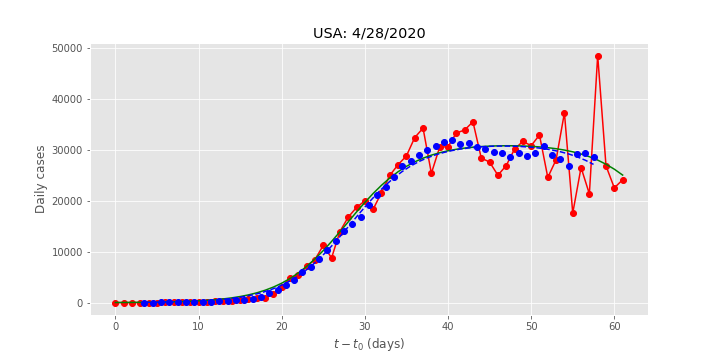}
\hspace{-0.04\textwidth}
\includegraphics[width=0.52\textwidth]{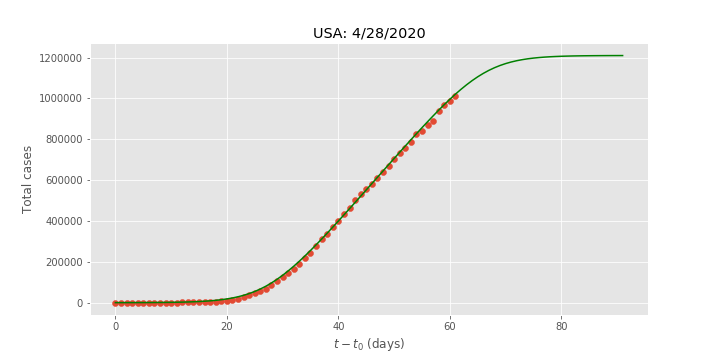}
\caption{
\small{
Data for China (CHN), Greece (GRC), USA and Spain (ESP) and the United States of America (USA).
Red dots are the daily data reported in Ref. \cite{ECDC_source}, the green lines are the fits by the FSIR model. 
The blue dots are seven-day running averages and the blue dashed lines are the fits by the FSIR model; 
in all cases the green and blue-dashed lines are essentially indistinguishable, except for USA near the 
end of the examined period.
}
}
\label{fig:data_outliers}
\end{figure}

In Fig. \ref{fig:data_outliers} we give some examples of the actual fits for the  
"outlier" countries (China, Greece, USA and Spain).
To have a measure of the fit that is comparable between different 
countries, we defined the ``quality of fit'' as:
\begin{equation}
Q_{\rm fit} = \frac{1}{N'} {\rm RMSE}
\label{eq:quality_fit}
\end{equation}
which is expressed as a percentage (multiplied by a factor of 100). 
The resulting values of the parameters, including our choices of $t_0$, 
are given in Table \ref{table:parameters}.

\begin{table}
\begin{center}
\begin{tabular}{| c |l r | r | r | r | r | r | r | r|}\hline
Index & Country & (Symbol) &  $\; \; \; \; t_0 $  & $\; \; t_1$ & $\Dt$ & $N'$  & $Q_{\rm fit}$ & $\NT$ &  $D$ \\ 
 &  &  & (days)  & (days) & (days) &  & (\%) &  & (days) \\  \hline 
0 & China & (CHN) & 17 & 18.5 & 11.1 & 7,343 & 16.19 & 81,100 & $-59$ \\
1 & Greece & (GRC) & 65 & 11.0 & 30.1 & 79 & 36.91 & 2,400 & $-4$ \\
2 & Australia & (AUS) & 61 & 23.8 & 5.7 & 1,143 & 6.61 & 6,500 & $-14$ \\
3 & Denmark & (DNK) & 75 & 9.63 & 32.0 & 258 & 21.30 & 8,200 & 7 \\
4 & France & (FRA) & 57 & 24.4 & 29.2 & 4,272 & 20.97 & 124,800 & 1 \\
5 & Germany & (DEU) & 57 & 22.6 & 29.7 & 5,246 & 17.22 & 156,000 & 0 \\
6 & Switzerland & (CHE) & 59 & 19.1 & 24.7 & 1,142 & 14.62 & 28,100  & $-6$ \\
7 & Italy & (ITA) & 53 & 19.0 & 42.8 & 4,774 & 15.13 & 204,000 & 3 \\
8 & United States & (USA) & 59 & 27.9 & 38.7 & 31,314 & 12.58 & 1,210,300 & 14 \\
9 & Spain & (ESP) & 56 & 24.3 & 28.7 & 7,417 & 14.42 & 212,800 & 0\\

\hline
\end{tabular}
\caption{
\label{table:parameters}
\small{
The values of the various parameters that enter in the FSIR model of Eq.~(\ref{eq:FSIR}),  
for the 10 countries considered (see text for details). 
The countries have been indexed according to their $N_T$ values, scaled by the 
population of each country. 
The next-to-last column includes the values for the {\em expected} total number 
of cases $N_T$ when the number of infections has dropped to near zero, and is an 
{\em extrapolated} value (rounded to the nearest 100).  The last column includes 
the number of days $D$ (counting from April 28) 
until the value of $N_T$ has reached 99\% of its final value;
a negative number (for China, Australia, Switzerland, and Greece) indicates that this date has already passed. 
}}
\end{center}
\end{table}

The values of the parameters obtained reveal interesting behavior.
\begin{itemize}
\item{$t_0$: 
The value of this parameter is similar for all countries, except for China
with $t_0 = 17$. 
This simply reflects the fact that the pandemic originated in China and then 
spread through the rest of the world. The rest of the countries have starting dates 
of the exponential increase within one week from the earliest, Italy (with $t_0 = 53$)
to the latest, such as Greece and Denmark (with $t_0 = 65$, and $75$, respectively).
The time lag between most countries and China is approximately 6 weeks. 
}
\item{$t_1$: 
This value indicates the position of the mid-point of the sigmoid  
representing the behavior of the susceptible population, S. 
The shorter it is, the sooner the country experiences the exponential increase
in the infected cases, thus urgently necessitating the introduction 
of health interventions and measures to limit the spread of the disease.  
The three countries with the shortest $t_1$ values 
are Greece, Denmark and China; unsurprisingly, these countries also have of the lowest 
number of cases per million, as shown in Fig. \ref{fig:daily_10}. 
}
\item{$\Delta t$: 
This value indicates the lag between the sigmoid that describes the recovered 
population ($\Rtild$) and the sigmoid of the susceptible population ($\Stild$). 
As such, it can be interpreted as the effective rate of removal ($\gamma$ in the SIR model).
 In Table \ref{table:parameters} we present the values of $\Delta t$ for each country. 
 The average of $\Delta t$ is close to 27.5 days ($\sim 4$ weeks), 
 a value consistent with a recently reported estimated median time of approximately 2 weeks 
 from onset to clinical recovery for mild cases, and 3--6 weeks for 
 patients with severe or critical disease (\cite{WHO_China_2020}, \cite{Zhou_2020}, \cite{Woelfel_2020}). 
Australia and China show an unusual low value, $\Delta t$= 5.7, and 11.1 days, respectively.
The value of this parameter has a significant effect on the total expected number of cases (see below). 
}
\item{$N'$: the value of this parameter is representative of the number of daily cases 
near the peak of the $\Itild$ curve. It is close to reported values for this quantity for 
all the countries.  Interestingly, if we assume that the total number of susceptible individuals is 
close to the population of each country, which in all cases is in the range of $N \sim 10^7 - 10^9$, 
then the ratio $n' = N'/N \to 0$, as we assumed for the FSIR model earlier.  \\
}
\end{itemize}

In Table \ref{table:parameters} we also include the values for the quality of the fit, 
which range from 6.6 (AUS) and 12.6 (USA) to 36.8 (Greece), representing a measure of the 
relative noise in the data; the noise is largest for Greece because the numbers are 
rather small. 
We have also considered fitting the FSIR model to seven-day running averages of the reported 
cases, and this in general makes almost no difference to the value of the parameters 
or the quality of the fit (see Fig. \ref{fig:data_outliers} for examples). 

Using the analytic expression for $\Itild(t)$ we can extrapolate to long times and 
try to obtain an estimate for the total value of cases over a long period, 
when the number of daily cases of infection have essentially dropped to 
negligible levels (this corresponds to $\Itild(t) \approx 0$). We call this asymptotic 
value $N_T$ and report it in Table \ref{table:parameters}.


\begin{figure}[h]
\hspace{-0.04\textwidth}
\includegraphics[width=0.52\textwidth]{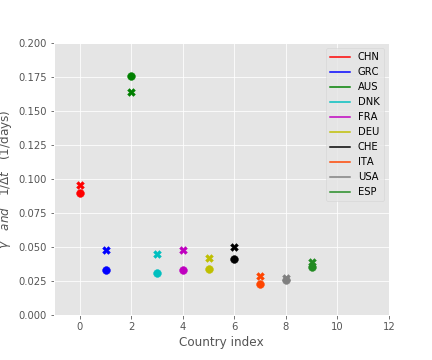}
\hspace{-0.04\textwidth}
\includegraphics[width=0.52\textwidth]{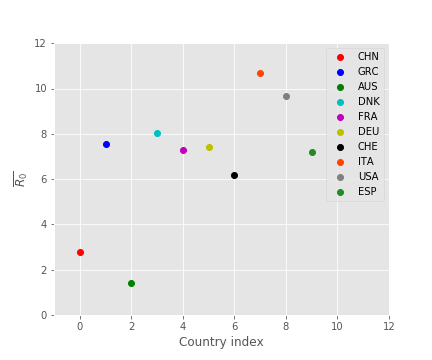}
\caption{
\small{
{\bf Left}: $ \overline{\gamma}$ and
$1/\Dt$ estimates for the 10 countries considered. The "filled x" symbols depict values of $\overline{\gamma}$, which is calculated as the time-average of the coefficient $\gamma (t)$ (Eq. (\ref{eq:gamma}), over the time period starting from $t_0$ 
until the expected total number of cases $N_T$ has reached 99\% of its final value. The filled circles depict values of $1/\Dt$. An explanation of the close proximity of the values of $\overline{\gamma}$ and $1/\Dt$ is presented in Section B (Analysis of the model). 
{\bf Right}: FSIR-estimated values for the reproduction number $\Rbar$ for the 10 countries considered. The larger the value $\Rbar$, the more adversely affected by the disease the country is.
}
}
\label{fig:former_Table_II}
\end{figure}

The average (over all countries in the set) of $\Dt$ is $27.31 \pm 11.35$ (median of $\Dt$ is 29.48), 
the average of $\overline{\gamma}$ is $0.058 \pm 0.042$,
and the average of $1/\Dt$ is $0.052 \pm 0.047$, (all numbers reported to 3 significant digits). 
It should be noted that the average of $\Dt$ over the set of the 50 countries 
mentioned earlier is $29.20 \pm 9.16$. 

The FSIR-estimated average of $\Delta t$ = $27.31 \pm 11.35$ 
yields an average reproduction number of $\Rbar \approx 6.83 \pm 2.84 $ 
(7.37, if we consider the median value of $\Dt$). 
Initial estimates of the early dynamics in Wuhan, China, 
suggested a value for $R_0$ in the range 2.2--2.7. 
For China, the FSIR model estimates $\Rbar = 2.77$. 
However, the FSIR estimates for the rest of the countries in the set, suggest much higher values of $\Rbar$. 
By conducting an elaborate analysis of datasets and data sources, estimating distributions of epidemiological parameters, 
and integrating uncertainties in parameters values, Sanche {\em et al.}\cite{Sanche_2020} 
reported a median $R_0 = 5.7$ (95\% CI 3.8 -- 8.9) for China. 
The FSIR estimated values of $\Rbar$ for the countries we study fall almost entirely within this range. 

Fig. \ref{fig:former_Table_II} depicts the FSIR-obtained values of $\overline{\gamma}$, plotted in conjuction with the $1/ \Dt$ values, for the 10 countries considered. For each country, the values are very close in magnitude, as was explained in a previous section. It should be noted that $\overline{\gamma}$ is 
calculated as the time average of the coefficient $\gamma (t)$, 
as presented in Eq. (\ref{eq:gamma}), over the time period starting from $t_0$ 
until the expected total number of infected people, $N_T$, has reached 99\% of its final value. Fig. \ref{fig:former_Table_II} also depicts the $\Rbar$ values for each country (calculated as $\Rbar \approx \alpha \Delta t $, where $\alpha \approx 0.25$ and the value of each country's $\Delta t$ is presented in Table \ref{table:parameters}).

As measured by the estimate of the basic reproduction number $\Rbar$, 
Italy is the country most adversely affected by the disease 
($\Rbar \approx 10.69 $), followed by the USA ($\Rbar \approx 9.78 $), 
Denmark ($\Rbar \approx 8.0$), Greece ($\Rbar \approx 7.53$), Germany ($\Rbar \approx 7.44 $), France ($\Rbar \approx 7.30 $), and Spain ($\Rbar \approx 7.17 $). 
Greece, although it suffered relatively small number of cases, has a large 
value for the basic reproduction number  ($\Rbar \approx 7.53$). 
USA, Spain, and Italy suffer the highest numbers of expected total cases 
($N_T$), whereas in the case of Greece the expected total number of cases is one of the lowest in the set, 
presumably due to the fast implementation of 
measures imposed by the government and followed by the citizens.
It should be noted that Greece has one of the smaller $t_1$ values. 
Australia, with $\Rbar \approx 1.42$, has the lowest value of the basic reproduction number, 
making the pandemic in this country to resemble the epidemiological characteristic of a seasonal flu. 


It is interesting to speculate on the meaning of these results. 
$N_T$, the total number of reported infections, is an important quantity 
because the number of fatalities (case fatality) is roughly proportional to this number, 
although the constant of proportionality varies in each country, ranging from 
a high of about 0.15 for Belgium, 0.14 for France, 0.13 for Italy, 
to a low of about 0.01 for Australia, and about 0.05 for Greece, China, and Denmark \cite{Johns_Hopkins_Mortality_Data}.

\begin{figure}[h]
\hspace{-0.04\textwidth}
\includegraphics[width=0.52\textwidth]{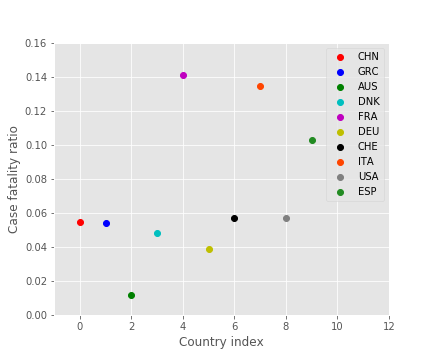}
\hspace{-0.04\textwidth}
\includegraphics[width=0.52\textwidth]{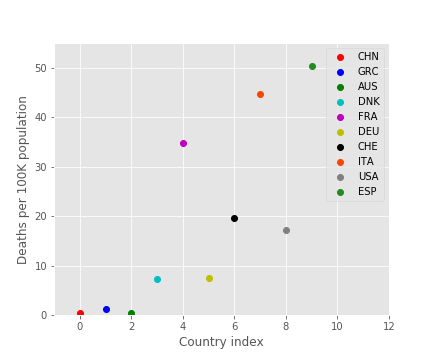}
\caption{
\small{
{\bf Left}: Case fatality ratios as reported in \cite{Johns_Hopkins_Mortality_Data} for 10 countries.
{\bf Right}: Deaths per 100K inhabitants for each country; 
these values closely follow the trends of the expected total number of infections scaled by the population, 
see Fig. \ref{fig:daily_10}. 
}
}
\label{fig:daily_11}
\end{figure}

Fig. \ref{fig:daily_11} presents the case fatality ratio and the deaths 
(COVID-related deaths) per 100K of the population, for each country. 
The case fatality ratio represents the mortality per absolute number of cases, that is,  
the total confirmed cases within a country. 
Greece, Denmark, and China have low values of case fatality 
ratios and deaths per 100K of the population, 
and so have Germany and USA. Australia has the lowest ratio. 
On the other hand, France, Italy, and Spain have the highest ratios.
In the scaled data, it is clear that Greece, China, and Australia
are atypical cases ("outliers") as having very low number of scaled deaths, 
while Spain, Italy, and France have the largest number of scaled deaths, 
in descending order; all sets represent deaths per 100K of the population of the respective country. 
Apparently, a low value of $t_1$ tends to imply a low $N_T$ 
and a low case fatality ratio and deaths, as the examples of China, Greece and Denmark demonstrate.  
In other words, early adoption of measures to contain the spread of the disease pay off.  
On the other extreme, the countries with high values of $t_1$ tend to have high values 
of $N_T$, especially Spain, 
which, along with Italy and France have very high case fatality ratios and deaths.

\pagebreak

\end{document}